\documentclass[structabstract]{aa}
\usepackage{graphicx}
\usepackage{txfonts}
\usepackage{aalongtable}
\usepackage{natbib}

\def\be{\begin{equation}}
\def\ee{\end{equation}}
\def\bea{\begin{eqnarray}}
\def\eea{\end{eqnarray}}

\newcommand{\sky}{{\mathrm{sky}}}

\newcommand{\plb}{{Phys.~Lett.~B}}

\begin{document}

\title{Testing \emph{CPT} Symmetry with CMB Measurements}

\author{Jun-Qing Xia\inst{1}
\and Hong Li\inst{2} \and Xiulian Wang\inst{3} \and Xinmin
Zhang\inst{1}}

\offprints{J.-Q. Xia, \email{xiajq@mail.ihep.ac.cn}}

\institute{Institute of High Energy Physics, Chinese Academy of
Science, P. O. Box 918-4, Beijing 100049, P. R. China. \and
Department of Astronomy, School of Physics, Peking University,
Beijing, 100871, P. R. China. \and Fakult\"{a}t f\"{u}r Physik,
Universit\"{a}t Bielefeld, D-33615 Bielefeld, Germany.}


\abstract {} {In this paper we study the possibility of testing
$CPT$ symmetry with Cosmic Microwave Background (CMB)
measurements.} {Working with an effective lagrangian of the photon
with $CPT$ violation ${\cal L} \sim p_{\mu}A_{\nu}\tilde
F^{\mu\nu}$ which causes the polarization vectors of the
propagating CMB photons rotated, we determine the rotation angle
$\Delta\alpha$ using the BOOMERanG 2003 and the WMAP3 angular
power spectra.} {In this analysis we have included the newly
released $TC$ and $GC$ ($l<450$) information of WMAP3 and found
$\Delta\alpha=-6.2\pm3.8$ deg at $68\%$ confidence level.} {This
result increases slightly the significance for the $CPT$ violation
obtained in our previous paper (Feng et al. 2006)
$\Delta\alpha=-6.0 \pm 4.0$ deg (1$\sigma$). Furthermore we
examine the constraint on the rotation angle from the simulated
polarization data with Planck precision. Our results show that the
future Planck measurement will be sensitive to $\Delta \alpha$ at
the level of $0.057$ deg and able to test the $CPT$ symmetry with
a higher precision.}


\maketitle


\section{Introduction}
\label{Introduction}

In the standard model of particle physics $CPT$ is a fundamental
symmetry. Probing its violation is an important way to search for
the new physics beyond the standard model. Up to now, $CPT$
symmetry has passed a number of high precision experimental tests
and no definite signal of its violation has been observed in the
laboratory. So, the present $CPT$ violating effects, if exist,
should be very small to be amenable to the experimental limits.

The $CPT$ symmetry could be dynamically violated in the expanding
universe \cite{Li:2007}. The cosmological $CPT$ violation
mechanism investigated in the literature
\cite{Li:2001st,Li:2002wd,Li:2004hh,Feng:2004mq,Li:2007} has an
interesting feature that the $CPT$ violating effects at present
time are too small to be detected by the laboratory experiments
but large enough in the early universe to account for the
generation of matter-antimatter asymmetry
\cite{Li:2001st,Li:2002wd,Li:2004hh,Li:2007}. And more
importantly, this type of $CPT$ violating effects could be
accumulated to be observable in the cosmological experiments
\cite{Feng:2004mq,Li:2007,Feng:2006dp}. With the accumulation of
high quality observational data, especially those from the CMB
experiments, cosmological observation becomes a powerful way to
test the $CPT$ symmetry.

Here we study the CMB polarizations and $CPT$ violation in the
photon sector with an effective lagrangian
\cite{Carroll:1989vb,Carroll:1990zs}:
\begin{equation}\label{Lagrangian}
\mathcal{L} = -\frac{1}{4}F_{\mu\nu}F^{\mu\nu}+\mathcal{L}_{cs}~,
\end{equation}
where $\mathcal{L}_{cs}\sim p_{\mu}A_{\nu}\tilde F^{\mu\nu}$ is a
Chern-Simons term, $p_{\mu}$ is an external vector and $\tilde
F^{\mu\nu}=(1/2)\epsilon^{\mu\nu\rho\sigma}F_{\rho\sigma}$ is the
dual of the electromagnetic tensor. This Lagrangian is not gauge
invariant, but the action is gauge independent if
$\partial_{\nu}p_{\mu}=\partial_{\mu}p_{\nu}$. This may be
possible if $p_{\mu}$ is constant in spacetime or the gradient of
a scalar field in the quintessential baryo-/leptogenesis
\cite{Li:2002wd,Li:2001st,quin_baryogenesis} or the gradient of a
function of the Ricci scalar in gravitational baryo-/leptogenesis
\cite{Li:2004hh,R}. The Chern-Simons term violates Lorentz and
$CPT$ symmetries when the background value of $p_{\mu}$ is
nonzero.

One of the physical consequences of the Chern-Simons term is the
rotation of the polarization direction of electromagnetic waves
propagating over large distances \cite{Carroll:1989vb}. From the
Lagrangian (\ref{Lagrangian}), we can directly obtain the equation
of motion for the electromagnetic field:
\begin{equation}\label{maxwell1}
\nabla_{\mu}(\nabla^{\mu}A^{\nu}-\nabla^{\nu}A^{\mu})=-p_{\mu}
\epsilon^{\mu\nu\rho\sigma}(\nabla_{\rho}A_{\sigma}-\nabla_{\sigma}A_{\rho})~.
\end{equation}
After imposing Lorentz gauge condition $\nabla_{\mu}A^{\mu}=0$, it
becomes:
\begin{equation}\label{maxwell2}
\nabla_{\mu}\nabla^{\mu}A^{\nu}+R^{\nu}_{\mu}A^{\mu}=-p_{\mu}
\epsilon^{\mu\nu\rho\sigma}(\nabla_{\rho}A_{\sigma}-\nabla_{\sigma}A_{\rho})~,
\end{equation}
where $R^{\nu}_{\mu}$ is the Ricci tensor. With the geometric
optics approximation, the solution to the equation of motion is
expected to be: $A^{\mu}={\rm Re}[(a^{\mu}+\epsilon
b^{\mu}+\epsilon^2 c^{\mu}+...)e^{iS/\epsilon}]$, where $\epsilon$
is a small number. With this ansatz, one can easily see that the
Lorentz gauge condition implies $k_{\mu}a^{\mu}=0$, where the wave
vector $k_{\mu}\equiv \nabla_{\mu}S$ is orthogonal to the surfaces
of constant phase and represents the direction which photons
travel along with. The vector $a^{\mu}$ is the product of a scalar
amplitude $A$ and a normalized polarization vector
$\varepsilon^{\mu}$, $a^{\mu}=A\varepsilon^{\mu}$, with
$\varepsilon_{\mu}\varepsilon^{\mu}=1$. Hence in the Lorentz
gauge, the wave vector $k_{\mu}$ is orthogonal to the polarization
vector $\varepsilon^{\mu}$. Substituting this solution into the
modified Maxwell equation (\ref{maxwell2}) and neglecting the
Ricci tensor we have at the leading order of $\epsilon$ the
equation is $k_{\mu}k^{\mu}=0$. It indicates that photons still
propagate along the null geodesics. The effect of Chern-Simons
term appears at the next order,
$k^{\mu}\nabla_{\mu}\varepsilon^{\nu}=-p_{\mu}\epsilon^{\mu\nu\rho\sigma}
k_{\rho}\varepsilon_{\sigma}$. We can see that the Chern-Simons
term makes $k^{\mu}\nabla_{\mu}\varepsilon^{\nu}$ not vanished.
This means that the polarization vector $\varepsilon^{\nu}$ is not
parallel transported along the light-ray. It rotates as the photon
propagates in spacetime.

We consider here the spacetime described by spatially flat
Friedmann-Robertson-Walker (FRW) metric. The null geodesics
equation is $(k^0)^2-k^ik^i=0$. We assume that photons propagate
along the positive direction of $x$ axis, \emph{i.e.}
$k^{\mu}=(k^0,k^1,0,0)$ and $k^1=k^0$. Gauge invariance guarantees
that the polarization vector of the photon has only two
independent components which are orthogonal to the propagating
direction. So, we are only interested in the changes of the
components of the polarization vector, $\varepsilon^2$ and
$\varepsilon^3$. Assuming $p_{\mu}=p_0$ to be a non-vanishing
constant, we obtain the following equations:
$d\varepsilon^2/d\lambda+\mathcal{H}k^0\varepsilon^2=p_0
k^0\varepsilon^3$,
$d\varepsilon^3/d\lambda+\mathcal{H}k^0\varepsilon^3= - p_0
k^0\varepsilon^2$, where we have defined the affine parameter
$\lambda$ which measures the distance along the light-ray,
$k^{\mu}\equiv dx^{\mu}/d\lambda$, and the reduced expansion rate
$\mathcal{H}\equiv \dot a/a$. The polarization angle is defined as
$\chi\equiv\arctan{(\varepsilon^3}/{\varepsilon^2})$. It is easy
to find that the rotation angle is
\begin{equation}\label{kmu1}
\Delta\chi\equiv \chi_0-\chi_z=-\int^{\eta_0}_{\eta_z} ~p_0
~d\eta=\int^{t_z}_{t_0} p_0 ~\frac{dt}{a}~,
\end{equation}
where the subscript $z$ is the redshift of the source when the
light was emitted. For CMB photons, the source is the last
scattering surface with $z\simeq 1100$ \footnote{Besides the CMB
photons which come from the last scattering surface, we might
observed the different CMB photons which travelled different
distances as well. However, we are only interested in linear
perturbations of CMB photons in this paper. CMB polarizations are
already linear phenomena. They are not existent at the zeroth
order. When calculating the variations of polarizations due to
$CPT$ violation in the perturbation theory up to linear order, we
may ignore the fluctuations of the travelling distances of CMB
photons. Otherwise, these fluctuations combined with polarizations
would give higher order result which are beyond the scope of this
paper. So, in this paper, we only consider linear perturbations
and we can assume that each CMB photon detected by us travelled
the same distance.}. The subscript $0$ indicates the present time.
As we know, a vector rotated by an angle $\Delta\chi$ in a fixed
coordinates frame is equivalent to a fixed vector observed in a
coordinates frame which is rotated by $-\Delta\chi$. So, with the
notion of coordinates frame rotation, the rotation angle is
\begin{equation}\label{kmu2}
\Delta\alpha=-\Delta\chi=\int^{t_0}_{t_z} ~ p_0 ~ \frac{dt}{a}~ =
p_0 r_z~.
\end{equation}
with $r_z$ being the comoving distance of the light source away
from us. This phenomena is known as ``cosmological birefringence".
This rotation angle $\Delta\alpha$ can be obtained by observing
polarized radiation from distant sources such as radio galaxies,
quasars and CMB.

The Stokes parameters $Q$ and $U$ of the CMB polarization can be
decomposed into a gradient-like ($G$) and a curl-like ($C$)
component \cite{Kamionkowski:1996ks}. For the standard theory of
CMB, the $TC$ and $GC$ cross-correlation power spectra vanish.
With the existence of cosmological birefringence, the polarization
vector of each photon is rotated by an angle $\Delta\alpha$, and
one would observe nonzero $TC$ and $GC$ correlations, even if they
are zero at the last scattering surface. Denoting the rotated
quantities with a prime, one gets \cite{Feng:2004mq,Lue:1998mq}:
\begin{eqnarray}\label{modify}
C_{l}^{'TC} &=& C_{l}^{TG}\sin(2\Delta\alpha)~, \nonumber\\
C_{l}^{'GC} &=&
\frac{1}{2}(C_{l}^{GG}-C_{l}^{CC})\sin(4\Delta\alpha)~,\nonumber\\
C_{l}^{'TG} &=& C_{l}^{TG}\cos(2\Delta\alpha)~,\nonumber\\
C_{l}^{'GG} &=& C_{l}^{GG}\cos^2(2\Delta\alpha) +
C_{l}^{CC}\sin^2(2\Delta\alpha)~,\nonumber\\
C_{l}^{'CC} &=& C_{l}^{CC}\cos^2(2\Delta\alpha) +
C_{l}^{GG}\sin^2(2\Delta\alpha)~,
\end{eqnarray}
while the temperature power spectrum $TT$ remains unchanged.


\section{Method and Results}
\label{results}

In our study we make a global analysis to the CMB data with the
public available Markov Chain Monte Carlo package
CosmoMC\footnote{http://cosmologist.info/.} \cite{Lewis:2002ah},
which has been modified to allow the rotation of the power spectra
discussed above, with a new free parameter $\Delta\alpha$. We
assume the purely adiabatic initial conditions and impose the
flatness condition motivated by inflation. Our most general
parameter space is: ${\bf P} \equiv (\omega_{b}, \omega_{c},
\Theta_{s}, \tau, n_{s}, \log[10^{10}A_{s}], r, \Delta\alpha)$,
where $\omega_{b}\equiv\Omega_{b}h^{2}$ and
$\omega_{c}\equiv\Omega_{c}h^{2}$ are the physical baryon and cold
dark matter densities relative to the critical density,
$\Theta_{s}$ is the ratio of the sound horizon to the angular
diameter distance at decoupling, $\tau$ is the optical depth to
re-ionization, $A_{s}$ and $n_{s}$ characterize the primordial
scalar power spectrum, $r$ is the tensor to scalar ratio of the
primordial spectrum. For the pivot of the primordial spectrum we
set $k_{s0}=0.05$Mpc$^{-1}$. In our calculation we have assumed
that the cosmic rotation angle is not too large and imposed a
conservative flat prior $-\pi/2\leq\Delta\alpha\leq\pi/2$.

\begin{figure}[t]
\begin{center}
\includegraphics[scale=0.5]{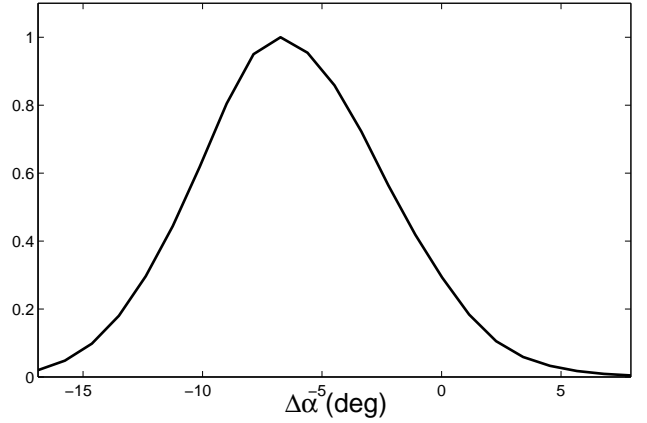}
\caption{One dimensional distribution on the rotation angle
$\Delta \alpha$ from WMAP3 and B03 data.} \label{fig1}
\end{center}
\end{figure}

In our calculation we combine the full data of WMAP3
\cite{WMAP31,WMAP32,WMAP33,WMAP34} (including the information of
$TC$ and $GC$ power spectra ($l<450$)) and BOOMERanG 2003 (B03)
\cite{B031,B032,B033}. We calculate the likelihood of $TT$, $TG$,
$GG$ and $CC$ power spectra using the routine for computing the
likelihood supplied by the WMAP team. As for the $TC$ and $GC$
power spectra data ($l<450$), which are just the preliminary
suboptimal results currently, we simply assume the Gaussian
likelihood function\footnote{We are very grateful to Professor
Eiichiro Komatsu (the member of WMAP group) for the email
communications on the use of the $TC$ and $GC$ data of WMAP3.
Because the current released $TC$ and $GC$ information of WMAP3
are not very accurate and do not include the correlations between
different $l$ values, our simple Gaussian likelihood function
Eq.(\ref{TBEB}) is accurate enough for the illustrative purpose.}:
\begin{eqnarray}\label{TBEB}
\mathcal{L}_{\rm TC/GC}&=&\exp
\left(-\frac{\chi^2_{\rm TC/GC}}{2}\right)~,\nonumber\\
\chi^2_{\rm TC/GC}&= &\sum_{\rm l}\left(\frac{C^{\rm TC/GC}_{\rm
l,th}-C^{\rm TC/GC}_{\rm l,obs}}{\sigma_{\rm l}}\right)^2~,
\end{eqnarray}
where $C^{\rm TC/GC}_{\rm l,th}$ and $C^{\rm TC/GC}_{\rm l,obs}$
denote the theoretical value and the experimental data of the $TC$
or $GC$ power spectra of WMAP3, $\sigma_{\rm l}$ denotes the
variance of estimated power spectra at each multi-pole.

\begin{figure*}[t]
\begin{center}
\includegraphics[scale=0.76]{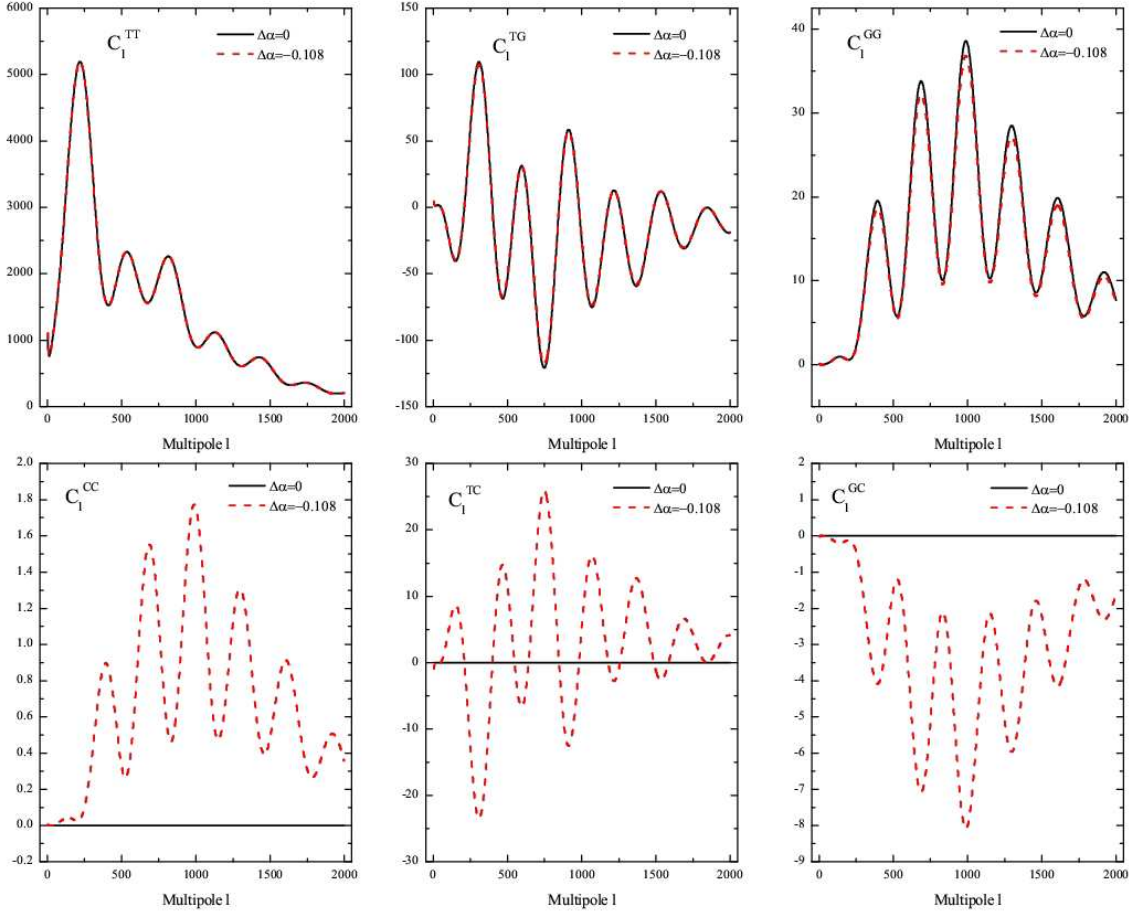}
\caption{The effects of the rotation angle $\Delta\alpha$ on the
power spectra of $TT$, $TG$, $GG$, $CC$, $TG$ and $GC$. The black
solid line is for the case of $\Delta\alpha=0$ and the red dashed
line is for $\Delta\alpha=-0.108$.} \label{fig2}
\end{center}
\end{figure*}

In the computation of the CMB spectra, we have considered lensing
contributions. Furthermore, we make use of the Hubble Space
Telescope (HST) measurement of the Hubble parameter $H_{0}\equiv
100$h~km~s$^{-1}$~Mpc$^{-1}$ by multiplying a Gaussian likelihood
function $h=0.72\pm0.08$ \cite{Hubble}. We also impose a weak
Gaussian prior on the baryon density
$\Omega_{b}h^{2}=0.022\pm0.002$ ($1\sigma$) from the Big Bang
Nucleosynthesis \cite{BBN}. Simultaneously we will also use a
cosmic age tophat prior as 10 Gyr $< t_0 <$ 20 Gyr.


In Figure \ref{fig1} we plot our one-dimensional constraint on the
rotation angle $\Delta\alpha$ from the combined WMAP3 and B03 data
and find that the current CMB polarization data favor a nonzero
rotation angle of the photons. The best fit value of the rotation
angle is $\Delta\alpha=-0.122=-7.0$ deg. Marginalizing the
posterior distributions, we find that the mean value of the
rotation angle is:
\begin{equation}\label{result}
\Delta\alpha=-0.108\pm0.067=-6.2\pm3.8~{\rm deg}~(1\sigma)~,
\end{equation}
which deviates the unrotated case $\Delta\alpha=0$ more than
$1\sigma$ and gives a weak evidence for $CPT$ violation. This
result is frequency independent, while the Faraday Rotation
\cite{Scannapieco:1997mt}, which also can give nonzero $TC$ and
$GC$ power spectra, is frequency dependent and of the much smaller
order than ours. On the other hand, currently several high
precision experimental tests have confirmed the $CPT$ symmetry and
do not detect apparent $CPT$ violation in the laboratory. However,
these two results are consistent. From equations
(\ref{kmu1},\ref{kmu2}) we can find that the size of the effects
of $CPT$ violation on CMB polarization power spectra is an
integrated effect and enhanced during the propagation time, while
the laboratory experiments just test the $CPT$ symmetry at
present. It means that the $CPT$ violating effect should be too
small to be detected at any specific time and enhanced during the
propagation time to be detectable by the CMB measurements.

In Figure \ref{fig2} we illustrate the effects of the rotation
angle $\Delta\alpha$ on the power spectra for two cases
$\Delta\alpha=0$ and $\Delta\alpha=-0.108$. The basic cosmological
parameters we choose for these plots are the best fit values of
global fitting: $\Omega_b h^2=0.0228$, $\Omega_m h^2=0.1320$,
$\tau=0.099$, $H_0=73.1$, $n_s=0.965$, $A_s=2.2\times10^{-9}$ and
$r=0.00866$. One can see from Figure 2 that the $C$-mode is very
sensitive to the rotation angle $\Delta\alpha$. Direct
measurements of the $TC$ and $GC$ power spectra are crucial and
would give more stringent constraints than ones of other power
spectra.

From the analysis of data points of $TC$ and $GC$ power spectra of
WMAP3 and B03, we can find that this negative rotation angle is
slightly preferred. Firstly, in the B03 data, the $TC$ power at
$l\sim250$ and $l\sim350$ are both negative, while it is positive
at $l\sim450$. The $GC$ power at $l\sim250$, $l\sim350$ and
$l\sim450$ are all negative. Based on the equations (\ref{modify})
and Figure \ref{fig2}, we can see that the $TC$ and $GC$ power
spectra of B03 really help to obtain this negative rotation angle.
On the other hand, the quality of $TC$ and $GC$ power spectra of
WMAP3 seems not very good. The data points are distributed around
the line $\Delta\alpha=0$ proportional and the error bars are very
large. Therefore, the WMAP3 polarization data can not give much
more contribution for testing the $CPT$ symmetry and not improve
the previous result of rotation angle significantly.

Recently Feng et al. (2006) used the WMAP3 and B03 polarization
data to constrain the rotation angle and found the similar result:
$\Delta\alpha=-6.0\pm4.0$ deg at $1\sigma$ confidence level. As we
show above, the $TC$ and $GC$ information are very important for
this type of analysis. In their analysis they did not include the
$TC$ and $GC$ information of WMAP3 and set the $TC$ and $GC$ of
power spectra of WMAP3 to be zero, $C^{\rm TC}_{\rm l}=0$ and
$C^{\rm GC}_{\rm l}=0$. This factor should affect the final
result. Comparing these two results, we can find that the mean
value of our result moves toward direction away from the zero and
the error bar shrinks a little, which strengthens the conclusion
on the $CPT$ violation.

Moreover, Cabella et al. (2007) perform a wavelet analysis of the
temperature and polarization maps of the CMB delivered by the
WMAP3 experiment which include the information of $TC$ and $GC$
power spectra. They set a limit on the CMB photon rotation angle
$\Delta\alpha=-2.5\pm3.0$ deg ($1\sigma$). They found no evidence
for the $CPT$ violation from WMAP3 polarization data. However, in
their analysis, they have not included the information of B03
polarization data. We also made an analysis without the B03 data
and found that $\Delta\alpha=-1.6\pm10.3$ deg ($1\sigma$) which
means that the WMAP3 data only did not give any significant
evidence for $CPT$ violation. Therefore, we believe that they
should obtain the similar result with ours if they also include
the B03 polarization data in their analysis.

\begin{table}\label{table}
TABLE I. Assumed experimental specifications. The noise parameters
$\Delta_T$ and $\Delta_P$ are given in units of $\mu$K-arcmin.
\begin{center}
\begin{tabular}{cccccc}
\hline \hline

$f_{\rm sky}$~ & ~$l_{\rm max}$~ & (GHz) &
~$\theta_{\rm fwhm}$~ & ~$\Delta_T$~~ & ~~$\Delta_P$~ \\

\hline

0.65 & 2500 & 100 & 9.5' & 6.8 & 10.9 \\
     &      & 143 & 7.1' & 6.0 & 11.4 \\
     &      & 217 & 5.0' & 13.1 & 26.7 \\

\hline \hline
\end{tabular}
\end{center}
\end{table}

Obviously the current CMB polarization data are not good enough to
verify this possible $CPT$ violation. We need more accurate CMB
data such as the Planck
measurement\footnote{http://sci.esa.int/science-e/www/area/index.cfm?fareaid=17/.}
in the near future. In Table I we list the assumed experimental
specifications of the future Planck measurement. The likelihood
function is $\mathcal{L}\propto \exp(-\chi_{\rm eff}^2/2)$ and
\begin{equation}\label{simu}
\chi^2_{\rm eff}=\sum_{\rm l}(2l+1)f_{\rm
sky}\left(\frac{A}{|\bar{C}|}+\ln\frac{|\bar{C}|}{|\hat{C}|}+3\right)~,
\end{equation}
where $f_{\sky}$ denotes the observed fraction of the sky in the
real experiments, $A$ is defined as:
\begin{eqnarray}\label{AAA}
A &=&
\hat{C}^{TT}_{l}(\bar{C}^{GG}_{l}\bar{C}^{CC}_{l}-(\bar{C}^{GC}_{l})^2)+\hat{C}^{TG}_{l}(\bar{C}^{TC}_{l}\bar{C}^{GC}_{l}-\bar{C}^{TG}_{l}\bar{C}^{CC}_{l})\nonumber\\
  &+& \hat{C}^{TC}_{l}(\bar{C}^{TG}_{l}\bar{C}^{GC}_{l}-\bar{C}^{TC}_{l}\bar{C}^{GG}_{l})+\hat{C}^{TG}_{l}(\bar{C}^{TC}_{l}\bar{C}^{GC}_{l}-\bar{C}^{TG}_{l}\bar{C}^{CC}_{l})\nonumber\\
  &+& \hat{C}^{GG}_{l}(\bar{C}^{TT}_{l}\bar{C}^{CC}_{l}-(\bar{C}^{TC}_{l})^2)+\hat{C}^{GC}_{l}(\bar{C}^{TG}_{l}\bar{C}^{TC}_{l}-\bar{C}^{TT}_{l}\bar{C}^{GC}_{l}) \nonumber\\
  &+& \hat{C}^{TC}_{l}(\bar{C}^{TG}_{l}\bar{C}^{GC}_{l}-\bar{C}^{GG}_{l}\bar{C}^{TC}_{l})+\hat{C}^{GC}_{l}(\bar{C}^{TG}_{l}\bar{C}^{TC}_{l}-\bar{C}^{TT}_{l}\bar{C}^{GC}_{l})\nonumber\\
  &+& \hat{C}^{CC}_{l}(\bar{C}^{TT}_{l}\bar{C}^{GG}_{l}-(\bar{C}^{TG}_{l})^2)~,
\end{eqnarray}
and $|\bar{C}|$ and $|\hat{C}|$ denote the determinants of the
theoretical and observed data covariance matrices respectively,
\begin{eqnarray}\label{CCC}
|\bar{C}|&=&\bar{C}^{TT}_{l}\bar{C}^{GG}_{l}\bar{C}^{CC}_{l}+2\bar{C}^{TG}_{l}\bar{C}^{TC}_{l}\bar{C}^{GC}_{l}
           -\bar{C}^{TT}_{l}(\bar{C}^{GC}_{l})^2\nonumber\\
         & &-\bar{C}^{GG}_{l}(\bar{C}^{TC}_{l})^2-\bar{C}^{CC}_{l}(\bar{C}^{TG}_{l})^2~,\nonumber\\
|\hat{C}|&=&\hat{C}^{TT}_{l}\hat{C}^{GG}_{l}\hat{C}^{CC}_{l}+2\hat{C}^{TG}_{l}\hat{C}^{TC}_{l}\hat{C}^{GC}_{l}
           -\hat{C}^{TT}_{l}(\hat{C}^{GC}_{l})^2\nonumber\\
         & &-\hat{C}^{GG}_{l}(\hat{C}^{TC}_{l})^2-\hat{C}^{CC}_{l}(\hat{C}^{TG}_{l})^2~.
\end{eqnarray}
The likelihood has been normalized with respect to the maximum
likelihood $\chi^2_{\rm eff}=0$, where $\bar{C}^{\rm XY}_{\rm
l}=\hat{C}^{\rm XY}_{\rm l}$. For more details of this calculation
we refer the readers to our previous companion paper
\cite{Xia:2007gz}. By the simulated data with Planck accuracy, we
find that the standard deviation of the rotation angle will be
significantly reduced to $\sigma=0.057$ deg. This result is much
more stringent than the current constraint. Assuming the mean
value unchanged, the cosmological $CPT$ violation will be
confirmed around $100\sigma$ confidence level with the future
Planck measurement.


In summary, probing the violation of fundamental symmetries is an
important way to search for the new physics beyond the standard
model. In this paper we have determined the rotation polarization
angle $\Delta\alpha$ with the CMB data from the WMAP3 and
BOOMERanG 2003. We find that a nonzero rotation angle of the
photons is mildly favored: $\Delta\alpha=-6.2\pm3.8$ deg
($1\sigma$). Our result shows a small violation of the $CPT$
symmetry, albeit not conclusive with the present CMB data. With
Planck data our simulations indicate that this type of $CPT$
violation could be confirmed significantly, or $CPT$ symmetry will
be verified with a high precision.


\begin{acknowledgements}

We acknowledge the use of the Legacy Archive for Microwave
Background Data Analysis (LAMBDA). Support for LAMBDA is provided
by the NASA Office of Space Science. We have performed our
numerical analysis on the Shanghai Supercomputer Center (SSC). We
are grateful to Eiichiro Komatsu for using of the $TC$ and $GC$
data of WMAP3. We thank Carlo Contaldi, Mingzhe Li and Hiranya
Peiris for helpful discussions. This work is supported in part by
National Natural Science Foundation of China under Grant Nos.
90303004, 10533010 and 10675136 and by the Chinese Academy of
Science under Grant No. KJCX3-SYW-N2.

\end{acknowledgements}


\end{document}